\documentclass[showpacs,twocolumn,nofootinbib]{revtex4}
\usepackage{amssymb}
\newcommand*\de{\mathrm{d}}
\renewcommand*\epsilon{\varepsilon}
\renewcommand*\phi{\varphi}
\renewcommand*\theta{\vartheta}
\begin{document}
\title{One-parameter teleparallel limit of Poincar\'e gravity} 
\author{ M. Leclerc}
\date{\today}
\affiliation{Section of Astrophysics and Astronomy, Department of Physics,
University of Athens, Greece}
\begin{abstract}
Poincar\'e gauge theories that, in the absence of spinning matter, 
reduce to the  
 one-parameter teleparallel theory are investigated with respect 
to their mathematical consistency and experimental viability. 
It is argued that the theories can be consistently coupled to the known 
standard model particles. Moreover, 
we establish the result that in the classical limit, such theories 
share  a large class of solutions with general relativity, containing, 
among others,  the four classical black hole solutions 
(Schwarzschild, Reisner-Nordstr\o m, 
Kerr and Kerr-Newman), as well as the complete class of 
Friedman-Robertson-Walker cosmological solutions, thereby extending 
 older viability results that were restricted to the correct Newtonian  
limit and to the existence of the Schwarzschild solution. 
\end{abstract}
\maketitle
\section{Introduction}

In the quest of a Poincar\'e gauge theory with a suitable limit  
 in the absence of spinning matter fields, we came to the 
 conclusion that no consistent theory with teleparallel limit exists 
if we require the field equations to reduce, in the spinless limit, 
to the field equations of general relativity. The reason for 
the inconsistency of such theories was found in the frame invariance 
of the field equations in the classical limit, i.e., invariance under 
a local Lorentz transformation of the tetrad field with the connection 
held fixed,  
thereby generalizing the analysis of Kopczy\'nski \cite{2} from 
purely teleparallel 
theories to the subclass of Poincar\'e gauge theories with teleparallel 
limit \cite{1}. (Whenever we use the 
expressions \textit{spinless} (or \textit{classical}) \textit{limit}, 
or simply \textit{limit}, we mean \textit{in the limit of vanishing 
spin}, i.e.,  in the absence of spinning matter fields.) 
On the other hand, we have already indicated in \cite{1} that one can 
modify the theories by adding a term of the form $\lambda T^{[ikl]} T_{[ikl]}$ 
to the Lagrangian (${}^{[...]}$ means total antisymmetrization), without 
sensibly 
changing the classical limit of the theory. The limit of such theories 
corresponds to the so-called \textit{one-parameter teleparallel theory}
 (sometimes called \textit{new general relativity} \cite{3}). 
As pointed out in \cite{2}, 
this reduces the frame invariance to a rather restricted class of 
transformations (special Lorentz transformations that leave the axial 
part of the torsion invariant), therefore taking care, to a certain extend, 
of the consistency problem. Moreover, it has already been established in 
 \cite{3} that the resulting field equations are still  in agreement with 
the experimental situation, independently of the value of $\lambda$. 
This result was based on the consideration of the Newtonian limit on one hand 
and on  the existence of the Schwarzschild solution on the other hand. 
This ensures  agreement with the classical experiments (perihelion shift, 
light deflection, red shift). 

Proceeding in the same fashion as in \cite{1}, 
the above results are easily generalized from 
the framework of purely teleparallel theories to the class of 
Poincar\'e gauge theories with a teleparallel limit. In this paper, 
we deal with two questions: 1) Are the theories with a teleparallel 
limit corresponding to the one parameter theory, $\lambda \neq 0$,  really 
consistent, or does the remaining frame invariance still represent a 
severe problem? 2) Can we extend the results on the experimental 
viability? Or otherwise stated, since general relativity  
seems to be in agreement 
with every experiment  
 carried out so far, does the one-parameter theory admit more 
general relativity solutions apart from the Schwarzschild metric?  
      
It should be noted that both questions can be dealt with  within the 
framework of the purely teleparallel gravity. The actual problems are thus 
unrelated to the fact that the theory  arises here as classical limit 
of a specific Poincar\'e gauge theory, whose \textit{main motivation} is, 
in the words of Kopczy\'nski, 
\textit{to grant the spin independent dynamical meaning} \cite{2}.     
In other words, the results are useful not only to those who, like us,  
consider the one-parameter theory as a starting point for the construction 
of viable Poincar\'e gauge theories with propagating curvature and torsion, 
but also to those who see in the theory merely an \textit{alternative} to 
general relativity. 

The paper is organized as follows. In section II, we introduce 
a concrete  Lagrangian,  exemplifying the theories we deal with. We derive 
the field equations and their classical limit. In section III
 we discuss the consistency question in  presence of 
  spinning matter fields, and finally, in section IV, we show that 
the classical limit of our theory, and more generally, of the one-parameter 
teleparallel theory, shares with general relativity  an important 
  class of metrics, 
containing, among many others, the known  black hole solutions and  
the relevant cosmological solutions. 

\section{Field equations and classical limit}

Our notations are identical to those of \cite{1}. Especially, the torsion 
is given by $T^a_{\ ik} = e^a_{k,i} + \Gamma^a_{\ bi}e^b_k - [i,k]$ 
and the curvature by $R^{a}_{\ bik} = \Gamma^{a}_{\ bk,i} + \Gamma^a_{\ ci}
\Gamma^c_{\ bk} - [i,k]$, where $e^a_i$ and $\Gamma^a_{\ bi}$ are the 
tetrad field and the Lorentz connection, respectively.    
As argued in \cite{1}, the suitable Lagrangian for the construction of 
Poincar\'e gauge theories with propagating torsion and curvature presenting 
a teleparallel limit is given by 
\begin{displaymath}
 L =  R - \frac{1}{4}T^{ikl}T_{ikl}-\frac{1}{2}T^{ikl}T_{lki} 
+ T^k_{\ ik}T^{mi}_{\ \ m}  + \lambda T^{[ikl]}T_{[ikl]},  
\end{displaymath}
where we have added 
the term $\lambda T^{[ikl]}T_{[ikl]}$,  
 in order to break the frame 
invariance. Clearly, this term can also be written as combination of 
terms of the form of the second and third terms in $L$, but we 
prefer here to write it separately because our study will focus on 
this term. (Note also that $L $ is a scalar, while the Lagrangian density 
is given by  $\mathcal L = e L$, where $e = \det e^a_i$.)  

Before we continue, let us remind that the consistency problems in 
the framework of the purely teleparallel theory 
(or \textit{new general  relativity}) has been considered 
by Kopczy\'nski \cite{2} some twenty years ago,  and in a series of 
follow up articles, notably in 
\cite{4}. A more complete set of references can be found in \cite{1}. 
Especially, in \cite{5}, the teleparallel theories were incorporated 
into the general framework of Poincar\'e gauge theory with the 
help of Lagrange multipliers. Recently, in \cite{6},  
the consistency problem has once again been discussed under the aspects 
of the symmetry properties of the stress-energy tensor. Such  continuous 
interest in the subject underlines its importance and  the need for a 
definite answer.  

If we proceed 
 adding any term of second order in the curvature tensor to $L$, we find 
a theory with the required properties. To be concrete, we consider the
following Lagrangian 
\begin{eqnarray}
 L&=&  R - \frac{1}{4}T^{ikl}T_{ikl}-\frac{1}{2}T^{ikl}T_{lki} 
+ T^k_{\ ik}T^{mi}_{\ \ m}  \nonumber \\ &&
+ \lambda T^{[ikl]}T_{[ikl]} 
+ a R^{ab}_{\ \ ik}R_{ab}^{\ \ ik} +  L_m. 
\end{eqnarray}
We have also included a matter Lagrangian $\mathcal L_m = e L_m$, 
(we write $L = L_0 + L_m $) and  define the canonical stress-energy 
tensor $T^a_{\ m} = (1/2e)
\delta \mathcal L_m / \delta  e^m_a $ as well as the spin density 
$\sigma_{ab}^{\ \ m} = (1/e) 
\delta \mathcal L_m / \delta \Gamma^{ab}_{\ \  m}$. 
Variation with respect 
to the independent fields $e^a_m $ and $\Gamma^{ab}_{\ \ m}$ leads to 
the following set of field equations 
\begin{equation}
\hat G_{ik} - \tau^{(1)}_{ik}   - \tau^{(2)}_{ik}   
  = T_{ik} 
\end{equation}
and
\begin{eqnarray}
  D_m R^{ablm}  
 +  \frac{\lambda}{a} K^{[abl]} &=&  
\frac{1}{4a} \ \sigma^{abl}.  
\end{eqnarray}
Here, $\tau^{(2)}_{ik} = -2a \left[ R^{ab}_{\ \ li}R_{ab\ k}^{\ \ l} 
- \frac{1}{4} g_{ik}R^{ab}_{\ \  lm}R_{ab}^{\ \ lm} \right ]$ 
is the symmetric stress-energy tensor of the (Lorentz)-Yang-Mills 
field, while $\tau_{ik}^{(1)}$ is given by 
\begin{eqnarray}
\tau_{ik}^{(1)} &=& 
\lambda \left[ -2 (T_{[kmi]}^{\ \ \ \ \ \ ;m}  - \frac{1}{2}
T_{[lmi]} K^{l\ m}_{\ k}) \right. \nonumber \\ &&  
\left. + (2 T_{[klm]}T^{ml}_{\ \ \ i}- \frac{1}{2}
g_{ik}T_{[lpm]}T^{lpm} ) \right].  
\end{eqnarray}

We refer to (2) and (3)  as Einstein and Cartan equation, respectively. 
Recall that we took the convention to denote with a hat all quantities 
constructed from the torsionless connection  and that $D_m$ is defined 
to act with $\Gamma^{ab}_{\ \ m}$ on tangent space indices $a,b,\dots$ and 
with $\hat \Gamma^i_{lm}$ (the Christoffel symbols) on spacetime indices 
$l,m,\dots$, while the semicolon denotes the usual covariant differentiation 
with the Christoffel connection. (It should be obvious what is meant by 
the somewhat sloppy, but convenient notation involving the total  
antisymmetrization over indices of different nature in the second 
term of Eq. (3).) 

The  case $\lambda = 0$ has been discussed in \cite{1}. It is an
 Einstein-Yang-Mills system for the Lorentz gauge field 
$\Gamma^{ab}_{\ \ m}$. The Einstein equation is symmetric in that case, 
as a result of the invariance of $\mathcal L_0$ under frame 
transformations, and thus, the coupling to spinning matter fields, 
which present an asymmetric stress-energy tensor, leads to inconsistencies. 

The scope of this article is to analyze the question whether the introduction 
of 
the term $\lambda T^{[ikl]}T_{[ikl]}$  
 resolves this problem (section III) 
and to what extend the classical limit is influenced by this modification
(section IV).   

In the classical, spinless limit, we see from Eq. (3) that the 
groundstate solution is 
given by $R^{ab}_{\ \ lm}=0 $, i.e., we have a teleparallel geometry. 
Further, we find the constraint on the torsion, $T^{[ikl]}= 0$ (no 
axial torsion). The remaining equations are thus
\begin{eqnarray}
\hat G_{ik} &=& T_{ik},  \\ 
T^{[ikl]} &=& 0,  
\end{eqnarray} 
under the condition $R^{ab}_{\ \ lm} = 0$. If we use the Poincar\'e gauge 
freedom (see \cite{1}) to choose $\Gamma^{ab}_{\ \ m} = 0$, the set of 
equations (5,6) essentially determines the tetrad field $e^a_i$. It is 
evident that the  same 
classical limit will be achieved for Lagrangians containing 
additional  terms quadratic in the curvature (e.g., $R^2$ or $R^{ik}R_{ik}$). 

It must be realized that our (classical) equations are stronger than 
those usually considered 
 in teleparallel theories, where the Lagrangian is considered 
right from the start to depend only on $e^a_i$ (\textit{tetrad gravity}). 
In the latter approach, Eq. (6) is absent and consequently, 
Eq. (5) includes  
the contribution from $\tau^{(1)}$. The statement  
 $\Gamma^{ab}_{\ \ m}$ is \textit{zero} (as a consequence of the field 
equations) is quite different from the statement 
$\Gamma^{ab}_{\ \ m}$ is \textit{absent} (right from 
the start). Both approaches share the solutions with $T^{[ikl]} = 0$. 

\section{Special frame transformations}

We now come to the problem that has been discussed extensively in 
\cite{1} for the case $\lambda = 0$. The fact that the Lagrangian 
$L_0$ is invariant under frame transformations leads to problems 
in the coupling to spinning matter fields, which are described by 
a non-frame invariant Lagrangian. This is most easily seen in the fact 
that the Einstein equation in such cases has a symmetric left hand side, 
while the stress-energy tensor of the matter fields will be asymmetric. 
We also showed that, even in the case where 
 $L_0$ is not frame invariant, but leads nevertheless 
to a frame invariant groundstate (by groundstate, we mean here the 
field configuration in the absence of spinning matter), problems arise 
when one considers spinning test particles entering the gravitational 
fields. It turns out that not all the measurable gravitational fields 
are determined by the field equations, and as a consequence, the behavior 
of  spinning test particles is not entirely predictable. 

Here, we wish to address the question whether those inconsistencies 
are remedied by the introduction of 
the term $\lambda T^{[ikl]}T_{[ikl]}$  
 in Eq. (1). 
The same problem, in the framework of the purely teleparallel theory, 
has been discussed in \cite{2} and in the follow-up articles (see the 
references in \cite{1}). 
Our specific Lagrangian is such that the (vacuum) field equations  
have the same symmetry properties (concerning the frame transformations) 
in both the general case [Eqs. (2,3)] and the classical limit [Eqs. (5,6)]. 
This allows us to make the discussion directly on the Lagrangian level. 
(If one adds terms like $R^2$ to $L_0$, the classical limit 
field configurations will have a larger symmetry than the Lagrangian itself, 
and the discussion will have to be transferred to test particles moving in 
 classical field configurations, as in \cite{1}. The conclusions, however, 
are the same in both cases.)  

First, note that under a frame 
transformation $e^a_i \rightarrow \Lambda^a_{\ b} e^b_i$ (with 
$\Gamma^{ab}_{\  \ i}$  fixed), 
the metric $g_{ik}$ and the curvature  $R^{ab}_{\ \ lm}$ are 
invariant, while the torsion transforms as 
\begin{equation}
T^a_{\  ik} \rightarrow \Lambda^a_{\ b} T^b_{\ ik} - e^b_{[k}
D_{i]}\Lambda^a_{\ b}, 
\end{equation}
or, if we consider the spacetime form $T^l_{\ ik} = e^l_a T^a_{\ ik}$, 
\begin{equation}
T^l_{\  ik} \rightarrow  T^l_{\ ik} - \Lambda_a^{\ c} e^l_c  e^b_{[k}
D_{i]}\Lambda^a_{\ b}.  
\end{equation}
Our notation is such that the transformation matrix $\Lambda^a_{\ b}$ can 
be dealt with as if it were a tensor, i.e., indices are raised and lowered 
with $\eta_{ab}$ and, in virtue of $\Lambda^c_{\ a} \Lambda^d_{\ b}\eta_{cd} 
= \eta_{ab}$, it holds $\Lambda^{\ a}_{b} = (\Lambda^{-1})^a_{\ b}$. 
(The meaning of the notation $D_i \Lambda^a_{\ b} $ is then also obvious.)
We see that in the absence of the second terms in (7,8), $T^a_{\ ik}$ 
transforms as a tensor (precisely, as a vector valued two-form), while 
$T^l_{\ ik}$ remains invariant. Note that a sufficient condition for this 
is not, as is sometimes stated, that $\Lambda^a_{\ b}$ is a rigid 
frame transformation ($\Lambda^a_{\ b,i}=0$), but rather that it is 
what could be called  covariantly rigid, namely  $D_i \Lambda^a_{\ b} = 0$. 
In the teleparallel 
limit $R^{ab}_{\ \ ik} =0$, we can choose $\Gamma^{ab}_{\  \ i} = 0$ and 
both conditions coincide. 

Apart from those covariantly rigid transformations, which do not represent 
a problem, since  the matter fields too are invariant under those, 
it is clear that $L_0$ (i.e.,  Lagrangian (1) without the part $L_m$) 
is  invariant under the specific subclass of frame transformations that 
leave invariant the totally antisymmetric part of the torsion, 
$T^{[ikl]}$. This is an immediate consequence of the fact that 
for $\lambda = 0$,  
$L_0$ is invariant under 
general frame transformations. That such restricted transformations 
exist has been established by Kopczy\'nski in \cite{2}, who has also 
given explicit examples.  

Clearly, for this subclass of restricted frame transformations, the whole 
discussion of \cite{1} 
applies, whenever the matter Lagrangian is not invariant under those specific 
transformations. We wish to point out in this article that  
no established theory exists that contains an elementary particle Lagrangian 
which is not invariant under the restricted frame transformations.  

First, consider bosonic fields. It is commonly accepted (see the review
article \cite{7}) that 
scalar fields couple only to the metric, and not to $\Gamma^{ab}_{\ \ i}$. 
The same holds for the Maxwell field and for gauge fields in general, 
which are considered to be  one-forms $A_i = A\wedge \de x^i$, and not 
vector valued scalars $A^a$. Any other coupling to the gravitational 
fields leads to a violation of the internal gauge invariance (see \cite{7}). 
Such fields, consequently, have a symmetric stress-energy tensor and a 
vanishing canonical spin density $\sigma_{ab}^{\ \ m} = (1/e) 
\delta \mathcal L_m / \delta \Gamma^{ab}_{\ \  m}$. It is also clear that 
nothing will  change on this situation if a symmetry is spontaneously broken, 
as in the case of the Weinberg-Salam model. Thus, although it is 
sometimes argued that massive vector fields (like the Proca field) can 
be coupled consistently to torsion (because they do not have the gauge 
invariance of their massless counterparts anyway), such models do not 
arise from the spontaneous breakdown of a gauge symmetry 
in a conventional Higgs model. Summarizing, the standard model 
 spin zero and spin one fields are described by frame invariant Lagrangians.  

Let us now turn to fermions. Committing a common abuse of language, fermions 
were actually what we meant by spinning particles in the first place, 
namely, particles with 
a non-vanishing spin density $\sigma_{ab}^{\ \ m}$. As it seems, all 
the particles so far observed (leptons and quarks) are spin 1/2 fermions 
described by a Dirac type Lagrangian. In virtue of the well known result 
that such particles, when minimally coupled to the connection 
$\Gamma^{ab}_{\  \ m}$, effectively couple only to the totally antisymmetric 
part of the torsion (see \cite{7}), their Lagrangian is 
invariant under the restricted frame transformations which leave the 
axial part of the torsion unaffected. Consequently, all physical quantities 
will not change during such a restricted frame change. Especially, 
the parts of the torsion that are not completely determined by the 
field equations (due to the freedom in the choice of a frame) do not 
couple to spin 1/2 test particles entering the gravitational fields and 
are thus physically irrelevant.  
 
This statement can already be found in \cite{3} and also in \cite{2}. 
It is based on the observation that the Lagrangian is constructed from 
a  suitable contraction of the 
quantities $\psi, D_m \psi $ and $e^a_m$, which, in the absence of 
torsion, transform homogeneously (i.e., as spinors and tensors, respectively)
under $e^a_m \rightarrow \Lambda^a_{\ b}e^b_m$, as we know from general 
relativity, and thus, the only non-homogeneously transforming quantity under 
the frame transformation is the torsion (appearing  in $D_m \psi$). 
Restricting the 
transformations to those leaving the only involved part of the torsion, 
$T^{[ikl]}$, invariant, leads to the above conclusion. 

We wish to point 
out here that, although the statement is correct in a certain sense, 
things are a little bit more involved, and the above line of argumentation 
is not correct. 

The Dirac Lagrangian, minimally coupled to the connection 
$\Gamma^{ab}_{\ \  m}$ is given by 
\begin{equation}
L_m = \frac{i}{2}(\bar \psi \gamma^m D_m \psi - \bar D_m \bar \psi \gamma^m 
\psi ), 
\end{equation}
where $D_m \psi = (\partial_m \psi - \frac{i}{4} 
\Gamma^{ab}_{\ \  m}\sigma_{ab}  ) \psi $, $\bar D_m \bar \psi = \partial_m 
\bar \psi + \frac{i}{4} \Gamma^{ab}_{\ \ m} \bar \psi \sigma_{ab} $, 
$\gamma^m = e^m_a \gamma^a$ and $\sigma_{ab} 
= \frac{i}{2} [\gamma_a, \gamma_b]$. 
The mass term $m \bar\psi \psi$ is unproblematic and has been  omitted 
for simplicity. 

The Lagrangian (9) 
 is invariant under the following gauge transformation 
\begin{eqnarray}
\psi & \rightarrow &  e^{-\frac{i}{4} \epsilon^{cd} \sigma_{cd}} 
\psi, \nonumber 
\\ \Gamma^a_{\ bm} & \rightarrow & 
\Lambda^a_{\ c} \Lambda_b^{\ d} \Gamma^c_{\  dm} 
- \Lambda_b^{\ c} \Lambda^a_{\ c,m}, \nonumber \\
e^a_m &\rightarrow &\Lambda^a_{\ b} e^b_m, 
\end{eqnarray}
under the condition that (infinitesimally) $\Lambda^a_{\ b} = \delta^a_b +
\epsilon^a_{\ b}$. We have identified this as (Lorentz part of the) 
Poincar\'e gauge transformation
in \cite{1}, in view of the relation of $e^a_m$ to the gauge potentials of the
translational part of the Poincar\'e group (see \cite{8}). The fact 
that (9) in invariant under (10) is not simply a result of suitable 
contractions on spinor and tensor indices, but in addition, a major 
role is played by the fact that the Lorentz generators $\sigma_{ab}$ 
do not commute with the Dirac matrix $\gamma^m = e^a_m \gamma^a$ that 
appears explicitely in (9). This is the result of the fact that, when 
dealing with the Lorentz group, as opposed to conventional gauge theories, 
\textit{inner} space, where the gauge transformation takes place, 
coincides with Dirac (or spinor) space. Therefore, in contrast to 
other gauge theories, it is not necessary to add an additional index 
to $\psi$. It carries already a Lorentz representation. 

Another, not unrelated remark concerns the transformation behavior of 
the Dirac matrices. Since the $\gamma^a$'s, just as $\psi$,  
are defined in Dirac space, they will transform 
as $\gamma^a \rightarrow S \gamma^a S^{-1}$, 
with $S = \exp [- (i/4) \epsilon^{cd}\sigma_{cd}]$. However, this 
transformation 
is exactly canceled by the transformation concerning the vector index, 
namely $\gamma^a \rightarrow \Lambda^a_{\ b} \gamma^b$, if, again, 
$\Lambda^a_{\ b} = \delta^a_b + \epsilon^a_{\ b}$. As a result, $\gamma^a$ 
and also $\sigma_{cd}$ are invariant under (10).  

In \cite{1}, we have split the transformation (10) in a pure Lorentz 
transformation, concerning only the Lorentz connection and leaving $e^a_m$ 
invariant, and a frame transformation, 
$e^a_m \rightarrow \Lambda^a_{\ b} e^b_m $, with $\Gamma^{ab}_{\ \ m}$ 
 invariant. Here, we deal with the question, whether (9) is invariant 
under the restricted class of frame transformations that leave $T^{[ikl]}$ 
invariant. Before we do this, it is necessary to define the behavior 
of the spinor field under those transformations. Following \cite{1}, 
we require again that the result of both transformations (Lorentz + frame)
should be equivalent to a Poincar\'e transformation (10). This leaves us 
essentially with two possibilities. The first is to extend the Lorentz 
transformation to 
\begin{eqnarray}
\psi & \rightarrow &  e^{-\frac{i}{4} \epsilon^{cd} \sigma_{cd}} 
\psi, \nonumber 
\\ \Gamma^a_{\ bm} & \rightarrow & 
\Lambda^a_{\ c} \Lambda_b^{\ d} \Gamma^c_{\  dm} 
- \Lambda_b^{\ c} \Lambda^a_{\ c,m}, \nonumber \\
e^a_m &\rightarrow &  e^a_m, 
\end{eqnarray} 
and consequently, the frame transformation affects only $e^a_m$ 
and leaves connection and spinor field invariant. 
However, it is 
quite easy to see that (9) is neither invariant under (11), nor 
under the corresponding frame transformation, not even under 
the restricted class. Indeed, even if we eliminate the torsion 
completely, (9) would still not be invariant under (11).

Nevertheless, it turns out that (9) is invariant under a different 
kind of restricted frame transformations. Let us  define, instead of 
(11), the following Lorentz transformation  
\begin{eqnarray}
\psi &\rightarrow & \psi, \nonumber 
\\ \Gamma^a_{\ bm} & \rightarrow & 
\Lambda^a_{\ c} \Lambda_b^{\ d} \Gamma^c_{\  dm} 
- \Lambda_b^{\ c} \Lambda^a_{\ c,m}, \nonumber \\
e^a_m &\rightarrow & e^a_m, 
\end{eqnarray}
as well as the corresponding frame transformation
\begin{eqnarray}
\psi & \rightarrow &  e^{-\frac{i}{4} \epsilon^{cd} \sigma_{cd}} 
\psi, \nonumber 
\\ \Gamma^a_{\ bm} & \rightarrow & \Gamma^a_{\ bm}, \nonumber \\
e^a_m &\rightarrow &\Lambda^a_{\ b} e^b_m.  
\end{eqnarray}
Clearly, the combination of (12) and (13) leads again to the Poincar\'e 
transformation (10), under which our Lagrangian is invariant. Therefore, 
invariance under (12) implies invariance under (13) and vice versa. 

Applying an infinitesimal transformation (12), $\delta \Gamma^{ab}_{ \ \ m} 
= - D_m \epsilon^{ab}$,  to the Lagrangian (9), 
we find for the variation 
\begin{equation}
\delta L_m = - \frac{1}{8} \bar \psi 
( \gamma^c \sigma_{ab} + \sigma_{ab} \gamma^c) e^m_c D_m \epsilon^{ab}.
\end{equation}
For general $\epsilon^{ab}$, this does certainly not vanish. However, for 
 our restricted class of transformations, we require the totally 
antisymmetric part of the torsion to be invariant. Therefore, from Eq. (8), 
we find  (for $\Lambda^a_{\ b} = \delta^a_b + \epsilon^a_{\ b}$) 
the constraint 
\begin{displaymath}
e_{[la} e^b_k D_{i]} \epsilon^a_{\ b}  = 0 
\end{displaymath}
where the antisymmetrization is to be performed over $[lki]$. Since the 
antisymmetrization process is not troubled by changing spacetime 
indices into tangent space indices,  we also have 
\begin{equation}
e^{m[c} D_m \epsilon^{ab]} = 0
\end{equation}
for the restricted class of transformations (antisymmetry over $[abc]$). 
It is well known that the factor 
$(\gamma_c \sigma_{ab} + \sigma_{ab} \gamma_c)$ appearing in (14) 
is totally antisymmetric and 
therefore, we can conclude that (9) is indeed invariant under (12) and 
(13) for the restricted class of transformations underlying the 
constraint (15). 

Unfortunately, (12) has nothing to do with a Lorentz gauge transformation, 
and in  (13), there are quite a few things more involved than just a 
\textit{frame transformation}. (Especially since the spinor space 
transformation on $\psi$ induces also a transformation of the Dirac 
matrices, as 
outlined above, which will have to be canceled by the corresponding 
Lorentz transformation.) 
 We do not know whether  the authors of   
\cite{2} and \cite{3} had the transformation (13) in mind, when they 
stated that the Dirac Lagrangian was invariant under a restricted class 
of frame transformations, but it is the only way to realize this. 

Fortunately, for the derivation of the field equations and of 
conservation equations, it is of no importance whether $\psi $ 
transforms, under a certain transformation, in the way of (11) or of (12), 
since in the total variation  $\delta \mathcal L_m  = (\delta \mathcal L_m
/ \delta e^a_m)  \delta e^a_m + (\delta \mathcal L_m / 
\delta \Gamma^{ab}_{\ \ m}) \delta \Gamma^{ab}_{\ \ m} 
+ (\delta \mathcal L_m / \delta \psi)  \delta \psi$, 
the last term does not contribute anyway, 
independently of the explicit form of $\delta \psi$, because on-shell, 
the Dirac equation $\delta \mathcal L_m / \delta \psi  = 0 $ will be 
satisfied. 
 
Although in a different way as naively expected, the spin 1/2 Lagrangian 
reveals itself to be invariant under those transformations   that leave 
the axial part of the torsion tensor invariant, and can thus 
be consistently coupled to our theory. 

Unfortunately, in the case of spin 3/2 particles, the situation is not 
so favorable. Those particles couple to each of the irreducible 
parts of the torsion (vector, axial and tensor), and a complete 
determination of the torsion by the field equations is thus required in 
order to get a predictable behavior of those particle. Otherwise stated, 
spin 3/2 particle Lagrangians are not invariant under the restricted 
frame invariance. However, it must be realized that such particles 
have not been observed in experiments so far and the reasons for 
introducing them are purely theoretical and confined to supersymmetric 
theories. More precisely, in supergravity theories, the supersymmetric 
partner of the spin two graviton is described by a spin 3/2  
field. However, we wish to point out that supergravity theories in 
their present form are not suitable for a generalization to theories 
with propagating torsion and curvature and therefore, they will 
eventually have to be adapted to those case, if we are to insist on our 
concepts of a similar description of spin and momentum  on one hand 
and curvature and torsion on the other hand. 

For simplicity, we confine ourselves to $N=1$ supergravity and refer 
the reader to the introductory review paper of 
van Nieuwenhuizen \cite{9}. The Lagrangian consists  essentially 
of the Einstein-Cartan term and a spin 3/2 gravitino term, if we 
neglect contributions that are due to the cosmological constant. 
However, it turns out that the action is not invariant under 
supersymmetry transformations unless constraints are introduced. 
In $N = 1$ supergravity, the constraint that is needed to render 
the action supersymmetric  coincides with the Cartan equation, which 
means essentially that the torsion is proportional to (a linear combination 
 of) the spin of the gravitino. (In more general theories, the constraints 
will not be given by a field equation and have to be put in by hand.)     
Clearly, a theory in which the torsion has to be eliminated in advance, either 
by solving the Cartan equation, or by fixing it introducing a constraint,  
is not suitable as a starting point for the 
construction of a theory with propagating torsion. Even in the case 
where the Cartan equation is still suitable as constraint, we would not be
able to solve it algebraically in generalized theories.         

Another point concerns the very base of the concept of supersymmetry. It 
is usually viewed as a gauge transformation that relates fermions to 
bosons and vice versa. This is indeed  realized in the $N=1$ model 
for instance, where the torsion is eliminated algebraically and 
the remaining dynamical fields are essentially the metric $g_{ik}$ (or 
the tetrad), and its superpartner, the spin 3/2 gravitino. However, 
if we start from a theory with propagating torsion and curvature, 
and try to construct its supersymmetric generalization, it can be 
expected that, apart from the gravitino, one  will also need 
a fermionic 
superpartner for  the connection $\Gamma^{ab}_{\ \ m}$, which then has the 
status of  an independent dynamical field. A similar view has been 
expressed in \cite{10} in the context of superconformal gravity. 

As a result, there is no much sense in analyzing the transformation 
behavior of conventional supergravity theories under frame transformations, 
since they are not suitable for the generalization to Lagrangians of the 
form (1) in their present form. The first step would be to construct 
supersymmetric theories allowing for both propagating torsion and curvature, 
and then to analyze the question whether their are theories among them 
that reduce,  in the spinless limit,  
to the one-parameter teleparallel field equations. 

We conclude that, as far as the standard model particles (gauge bosons, 
leptons, quarks, scalar fields) are concerned, the restricted frame invariance 
of the gravitational Lagrangian (1) does not lead to inconsistencies, 
because those parts of the torsion that remain undetermined by the 
field equations, as a result of the frame invariance, do not couple 
to any of those particles and are thus unobservable. This is a result of 
the fact that the corresponding particle Lagrangians are invariant too 
under the restricted class of frame transformations.

\section{The conformal Kerr-Schild class}

This section is entirely devoted to the classical limit of the field equations
  and 
to the problem of the experimental viability of the theory. (Unfortunately, 
experiments involving spin effects of elementary particles in connection 
with gravity have not been carried out to date, because of the smallness 
of the expected effects.) 

We take here the standpoint that general relativity is experimentally viable 
in all aspects, and therefore, instead of confronting  our theory directly  
with the experiment, we simply compare it with general relativity. 

As we have pointed out at the end of section II, our equations (5,6) 
are not exactly those of the purely teleparallel theory, because in 
the latter, the Cartan equation is naturally absent. They share however 
the class of solutions with $T^{[ikl]} =0$. Clearly, in virtue of Eq. (5), 
every solution of our classical equations is also a solution of  
general relativity.  
 
Therefore, we have already solved our problem: Every solution of (5,6) 
is also a solution of general relativity (and a solution of the 
conventional one-parameter teleparallel theory). General relativity and 
our theory in the classical limit have in common the solutions that satisfy  
 $T^{[ikl]} = 0$. 

More precisely, one should say \textit{the solutions that can be brought into  
a form where $T^{[ikl]}= 0$}, since to a given metric, there are many 
possible choices for  
 the tetrad field (related by frame transformations), 
leading to different torsion tensors.  (Recall that in the classical limit, 
we can always choose $\Gamma^{ab}_{\ \ m} = 0$.)  

An example of such a class of metrics  can already be found  in \cite{3}. 
It is the class of metrics that are diagonal and contains, according to the 
authors,  many static solutions with high symmetry, especially the 
Schwarzschild and the Reisner-Nordstr\o m solutions. 

We have found yet another, larger class of metrics, namely the 
conformal Kerr-Schild class, given by metrics of the form 
\begin{equation} 
g_{ik} = \phi^2(\eta_{ik} - k_i k_k), 
\end{equation}
where $\phi$ is a scalar field, $\eta_{ik}$ the Minkowski metric and 
$k_i $ a null-vector field, satisfying $\eta^{ik} k_i k_k = 0$. 

For simplicity, we first investigate the case $\phi = 1$. If we define 
$k^i = \eta^{ik} k_k$, then we have for the inverse of the metric 
$g^{ik} = \eta^{ik} + k^i k^k$ and consistently, $k^i = g^{ik}k_k$. 
Note also the relations $g_{ik} k^i k^k = \eta_{ik}k^i k^k = 0$. 
We choose the tetrad field as 
\begin{equation}
e^a_i = \delta^a_i - \frac{1}{2} \delta^a_l k^l k_i. 
\end{equation}
Obviously, $e^a_i e^b_k \eta_{ab} = \eta_{ik} - k_i k_k$. The inverse 
is given by $e^i_a = \delta^i_a + \frac{1}{2} \delta^l_a k^i k_l$. 
For the torsion, we find 
\begin{eqnarray}
T^a_{\ ik} &=& e^a_{k,i}- e^a_{i,k} \nonumber \\
&=& -\frac{1}{2} \delta^a_m (k^m k_k)_{,i} 
+ \frac{1}{2} \delta^a_m (k^m k_i)_{,k},   
\end{eqnarray}
and its  spacetime form reads 
\begin{equation}
T_{mik} = -\frac{1}{2} (k_m k_k)_{,i} + \frac{1}{2} (k_m k_i)_{,k}. 
\end{equation}
In deriving those equations, it is helpful to note the relation 
$k^i k_{i,k} = k_i k^i_{\ ,k} =  0$.  
Obviously, the totally antisymmetric part of $T_{mik}$ vanishes. 

In order to generalize this result to the metrics of the form (16), we 
show the transformation behavior of $T^a_{\ ik}$ under a conformal 
transformation $e^a_m \rightarrow \phi e^a_m $, i.e., $g_{ik} \rightarrow 
\phi^2 g_{ik}$. We find for the transformed torsion 
\begin{equation}
\tilde T^a_{\ ik} =  \phi T^a_{\ ik} + \phi_{,i}\ e^a_k - \phi_{,k}\ e^a_i, 
\end{equation}
leading to 
\begin{equation}
\tilde T_{mik} = \phi^2 T_{mik} + \phi [ \phi_{,i}\ g_{mk} - \phi_{,k}\ 
 g_{mi}].
\end{equation}
Clearly, if the axial  part of $T_{mik}$ vanishes, then so 
does the axial part of $\tilde T_{mik}$, which completes  our argument.  

Let us point out that the above transformation of $T^a_{\ ik}$ follows 
directly from the specific form of the torsion tensor in the teleparallel 
limit ($\Gamma^{ab}_{\ \ i} = 0$) of our theory, as given by the 
first line in (18), and the transformation behavior of $e^a_m$. 
This is not to be confused with what is usually called \textit{conformal
transformation} in the framework of Poincar\'e gauge theory, which requires 
the specification  
 of the transformation behavior of 
both tetrad  and  connection. Such
transformations have been considered in, e.g., \cite{11}-\cite{13}
in relation with Riemann-Cartan-Weyl geometries. A complete 
classification of conformal, projective and dilational (or scale) 
transformations in the more general framework of metric affine 
theory can be found in \cite{7}. 

It is important to remark that, although the explicit form of the metric 
(16) is highly coordinate dependent, the final relation, $T^{[ikl]} = 0$, 
is a tensor relation, and therefore holds independently of the coordinate 
choice. This means that, whenever a solution of general relativity 
can be brought into the form (16),  there will always be 
a choice for the tetrad, unique up to the restricted frame transformations  
 underlying the constraint (15), such that $T^{[ikl]}$ vanishes. This 
particular tetrad is then a solution to our classical field equations 
(5,6), and more generally to the field equations of the one-parameter 
teleparallel theory.  

In the case where such a general relativity solution is given in a 
different coordinate system, we do not know a priori what will be 
the corresponding tetrad field satisfying $T^{[ikl]} =0$, but 
 we certainly know that it exists. For instance, to the Schwarzschild 
solution in its conventional, diagonal, form corresponds a diagonal 
tetrad field (see \cite{3}). 

The class (16) corresponds to the metrics conformal to the Kerr-Schild 
class and is well known in the context of general relativity (see, e.g.,  
 \cite{14}).  In the simple Kerr-Schild class ($\phi = 1$), 
we find, among others,  
the known black hole solutions (Schwarzschild, Reisner-Nordstr\o m, 
Kerr and Kerr-Newman) as well as exact wave solutions (see \cite{14}). 
On the other hand, for $k_i = 0$, we are dealing with conformally flat 
metrics. This class contains, for instance, the important cosmological   
Friedman-Robertson-Walker solutions, when transformed into a suitable 
coordinate system. Examples of solutions containing both $\phi$ and 
$k_i$ can be found in \cite{15}, where inner Schwarzschild solutions 
have been derived in the form (16). 

Consequently, our classical limit field equations (5,6), although 
stronger than the equations of general relativity, nevertheless 
admit  a large class of solutions, containing all those  solutions that 
have been related to experiments so far. The theory described by 
the Lagrangian (1)  can therefore be considered to be experimentally 
viable. 

Let us note once again that in the conventional one-parameter 
teleparallel theory, as conceived by \cite{3}, 
the constraint $T^{[ikl]}$ is missing, and the Einstein equations 
therefore contain additional contributions. This allows eventually 
for more solutions, but does not affect the above arguments. 

It is interesting to note that a teleparallel form 
of the Kerr metric has also be found for the case $\lambda = 0$ (i.e., 
the teleparallel equivalent of general relativity)
in \cite{16}, where 
the torsion is, in complete contrast to our case, purely axial. 
The authors then conclude, correctly,  on a spin precession 
effect for a Dirac electron in this field. On the other hand, 
using the Kerr solution in the form (17), we will find no such effect.  
This is a concrete example for the inconsistency of the 
theory with $\lambda = 0$, which allows for both, 
physically inequivalent solutions, illustrating very clearly the 
arguments given in \cite{1}. Once we choose $\lambda \neq 0$, 
the only allowed solutions (for a classical source) are constraint 
to $T^{[ikl]}  = 0$, and the remaining freedom in the choice  
of the tetrad field (restricted frame invariance) does not affect the 
spin precession. 

It is also worthwhile to note that, in view of Eq. (6) and the results of 
sec. III, all test particles in a classical field configuration 
will move on  geodesics, since the
the only part of the torsion they eventually couple to is zero. 
More generally, the evolution of a spinning particle (momentum propagation, 
spin precession, \dots) does not differ from that of a spinless particle. 
Consequently, in order to study intrinsic 
spin effects, one has to consider fields 
created by spinning sources. 

Finally, we wish to point out that in the framework of the purely teleparallel 
theory, the Kerr black hole counterpart of the one-parameter theory 
has been found (see \cite{17} and \cite{18}) 
and later on generalized to the charged case 
\cite{19}. The solutions contain an additional parameter (apart from 
the angular momentum parameter of the Kerr metric and the charge parameter 
in the Kerr-Newman case), and coincide with the Kerr and Kerr-Newman 
solutions only for a specific value of that parameter. 
It is concluded that no experimental distinction of 
  those solutions to the corresponding general relativity solutions 
 is possible in the presence of scalar, spin 1/2 and 
Yang-Mills fields. This is consistent with our results of section III. 
Indeed, the solutions corresponding to different values of the parameter 
 only differ by their axial torsion parts, which do, however, 
not couple to the mentioned fields.

\section{Final remarks}

Hoping to resolve the inconsistency of the Poincar\'e gauge theories 
with a teleparallel limit, we followed  the step performed  by Hayashi and 
Shirafuji in the framework of the purely teleparallel theory and 
introduced the term $\lambda T^{[ikl]} T_{[ikl]}$ 
that breaks 
the frame invariance of the theory. We showed that this improved 
 theory can indeed be coupled consistently to all the 
standard model  elementary particles, using the fact that the 
remaining, restricted, frame invariance of the gravitational Lagrangian 
is also a symmetry of the matter  Lagrangians (Yang-Mills, Higgs, 
Dirac \dots). It turns out that in the spin 1/2 case, in order to assure  
this  invariance, a spinor transformation has to be performed 
simultaneously  
 with the frame change. 

In a second part, we concentrated on the 
experimental aspects of the theory. Although the field equations, in the 
classical limit,  now differ from those of general relativity  
(in contrast to the theories discussed in \cite{1}), we were able to 
show that they share a very important class of solutions with  
the latter. More precisely, every solution of general relativity 
 that can be brought into 
a form conformal to the Kerr-Schild metric is also a solution to 
our theory. This class contains every solution so far involved in 
any experimental procedure, and many other solutions that have been 
subject to theoretical investigations in the past, notably the 
black hole solutions (Schwarzschild, Kerr, Reisner-Nordstr\o m, Kerr-Newman)
and many cosmological solutions. Thus, we can conclude that, 
as far as the classical limit is concerned, the theory is experimentally 
indistinguishable from general relativity. Experiments involving 
the gravitational interaction of particles with intrinsic spin are 
needed to conclude further. 

In spite of these positive results, we would like to express 
some doubts that the introduction of the term $\lambda T^{[ikl]} T_{[ikl]}$ 
 is really 
the way to proceed. Concerning the free gravitational fields, 
it seems quite arbitrary to break the frame invariance 
\textit{as far as possible}, 
leading to constraints in the field equations 
that leave \textit{just enough} freedom to allow for the most important 
general relativity solutions. The fact that this (explicit) 
symmetry breaking reveals itself as, again,  \textit{just enough} to 
resolve the inconsistencies concerning the coupling to elementary 
particles could be seen as a coincidence. Or, otherwise stated, does 
the fact that, so far, the only observed half-integer spin particles are 
of spin 1/2,  justify the use of a theory that excludes the coupling to 
higher spin particles a priori?  

On the other hand, apart from the inconsistencies arising from the 
restricted frame invariance, there are also algebraic consistency 
problems. Indeed, it is well known that in a Riemannian geometry, 
the presence of matter fields with spin higher than 2 leads to 
severe restrictions (the so-called consistency conditions) on the 
curvature tensor which are so strong that it is actually almost impossible 
to construct reasonable theories involving such fields in curved spacetime. 
Extending the analysis to the case of Riemann-Cartan geometry, it 
was found that strong consistency conditions on curvature and torsion arise 
already for fields with spin larger than 1/2. (Except for spin one 
gauge fields, which are considered to be one-forms and do not couple to 
torsion.) For details, see \cite{20}-\cite{23} and references therein. 
Thus, in view of these algebraic inconsistencies concerning higher spin 
fermions on a Riemann-Cartan background geometry, it does not seem 
much of a drawback to have additional dynamical inconsistencies 
arising as a result of the specific choice of the gravitational 
Lagrangian. It simply means that we cannot allow for 
higher spin fermions. (Note that supergravity, as a specific combination 
of spin 2 and spin 3/2 fields, is not subject to the above problem, but as 
mentioned earlier, the torsion is non-dynamical and directly related to 
the spin density of the gravitino. Therefore, in that framework, 
it makes hardly sense to talk of a gravitino in a torsion \textit{background}.)

Another point concerns the interpretation of the restricted frame 
invariance, especially in view of the spinor transformation involved 
[see Eqs. (12,13)]. Even in the unrestricted form, the transformations 
 cannot be identified neither as gauge, nor as frame transformations 
in the strict sense. The role of the additional constraint [Eq. (15)] 
is even less clearer. One might prefer to work with theories 
that are free of  symmetries to which no  physical interpretation
can be asserted to, 
as is the case for Einstein-Cartan theory or other Poincar\'e gauge 
models with a Riemannian (instead of teleparallel) limit. There, 
however, one has to pay the price of either  putting  some limits 
on the involved coupling constants (in order to reduce the deviation 
from general relativity to an experimentally acceptable level) or 
  of having only non-propagating torsion fields (Einstein-Cartan). 

Finally, and most importantly, there is the question of how we 
deal with compound particles, or macroscopic, spin polarized bodies
(e.g., neutron stars). It has been argued in \cite{24} that in 
the limit of large spin ($>>1$),  particles will couple to 
the complete torsion tensor, and that this will be the case for 
macroscopic spin polarized matter too (see \cite{25} for a detailed 
review concerning the spin-torsion couplings and the resulting equations 
of motion). A concrete example is given by 
 the model of the Weyssenhoff 
fluid with intrinsic spin,  where the spin density is considered to be of  
the form $\int e \sigma_{ab}^{\ \ m} 
\de^3 x \sim \sigma_{ab} u^m$, where $\sigma_{ab}$ is the spin tensor and 
$u^m$ the four-velocity 
 (see \cite{26} for the detailed treatment of the Weyssenhoff fluid in a
 Riemann-Cartan framework). Especially, since $\sigma_{kli}$ is not 
totally antisymmetric, it will couple to non-axial torsion contributions.    
Thus, if we do not want to run into trouble, we have to reconsider 
those  semi-classical models of intrinsic spinning matter distributions. 

We conclude that the theories presented in this article are consistent  
in the presence of the standard model matter fields and are experimentally 
viable. The coupling to higher half-integer spin fields is not possible 
and there are doubts on the consistency of the macroscopic spin-torsion 
coupling. Further, the restricted frame invariance of the theories 
lacks a physical interpretation and the introduction of the term 
 $\lambda T^{[ikl]}T_{[ikl]}$  seems arbitrary.  

\section*{Acknowledgments}

This work has been supported by EPEAEK II in the framework of ``PYTHAGORAS 
II - SUPPORT OF RESEARCH GROUPS IN UNIVERSITIES'' (funding: 75\% ESF - 25\% 
National Funds).


\begin{thebibliography}{10}

\bibitem{1} M. Leclerc, 
Phys. Rev. D {\bf 71}, 027503 (2005) 

\bibitem{2} W. Kopczynski, 
J. Phys. A {\bf 15}, 493 (1982)

\bibitem{3} K. Hayashi and T. Shirafuji, 
Phys. Rev. D {\bf 19}, 3524 (1979)

\bibitem{4} J.M. Nester, Class. Quant. Grav. {\bf 5}, 1003 (1988) 

\bibitem{5} Y.N. Obukhov, and J.G. Pereira, Phys. Rev. D {\bf 67}, 
044016 (2003)

\bibitem{6} Y.N. Obukhov, and J.G. Pereira, Phys. Rev. D {\bf 69}, 
128502 (2004)

\bibitem{7} F.W. Hehl, J.D. McCrea, E.W. Mielke, and Y. Ne'eman, 
Phys. Rep. {\bf 258}, 1 (1995)  

\bibitem{8} R. Tresguerres and E.W. Mielke, Phys. Rev. D {\bf 62}, 044004 
(2000)

\bibitem{9}  P. van Nieuwenhuizen,  arXiv:hep-th/0408137
 
\bibitem{10} P.K. Townsend and P. van Nieuwenhuizen, 
Phys. Rev. D {\bf 19}, 3166 (1979)

\bibitem{11} T. Dereli, and R.W. Tucker, Phys. Lett. B {\bf 110}, 206 (1982)

\bibitem{12} H.T. Nieh, Phys. Lett. A {\bf 88}, 388 (1982) 

\bibitem{13} Y.N. Obukhov, Phys. Lett. A {\bf 90}, 13 (1982) 


\bibitem{14} 
H. Stephani, D. Kramer, M. MacCallum, C. Hoenselaers, and  E. Herlt, 
\textit{ Exact solutions of Einstein's field equations}, 
2nd ed., Cambridge University Press, 2003  


\bibitem{15} N. Dadhich, Gen. Rel. Grav. {\bf 28}, 1455 (1996) 

\bibitem{16} J.G. Pereira, T. Vargas, and C.M. Zhang, 
Class. Quant. Grav. {\bf 18}, 833 (2001)


\bibitem{17} N. Toma, Prog. Theor. Phys. {\bf 86}, 659 (1991)

\bibitem{18} R.D. Hecht, Phys. Lett. A {\bf 165}, 194 (1992) 

\bibitem{19} T. Kawai, and N. Toma, Prog. Theor. Phys. {\bf 87}, 583 (1992)


\bibitem{20} T. Kimura, J. Phys. A {\bf 14}, L329 (1981) 
\bibitem{21} N.H. Barth, and S.M. Christensen, J. Phys. A {\bf 16}, 543 (1983)
\bibitem{22} W.H. Goldthorpe, Nucl. Phys. B {\bf 170}, 307 (1980)
\bibitem{23} Y.N. Obukhov, J. Phys. A {\bf 16}, 3795 (1983) 


\bibitem{24} K. Hayashi, K. Nomura, and T. Shirafuji, 
Prog. Theor. Phys. {\bf 86}, 1239; {\bf 87}, 1275 (1992)  

\bibitem{25} M. Leclerc, 
Class. Quant. Grav. {\bf 22}, 3203 (2005) 

\bibitem{26} Y.N. Obukhov, and V.A. Korotkii, 
Class. Quant. Grav. {\bf 4}, 1633 (1987) 







\end{thebibliography}
\end{document}